\documentclass[epj]{svjour}
\usepackage{graphics}
 
\begin{document} 
 
\newcommand{\p}{\partial} 
\newcommand{\ls}{\left(} 
\newcommand{\rs}{\right)} 
\newcommand{\beq}{\begin{equation}} 
\newcommand{\eeq}{\end{equation}} 
\newcommand{\beqa}{\begin{eqnarray}} 
\newcommand{\eeqa}{\end{eqnarray}}

\title{Fluctuations of fragment observables
} 

\author{F. Gulminelli\inst{1} \and M.D'Agostino\inst{2}}
\institute{LPC Caen (IN2P3-CNRS/Ensicaen et Universit\'e),
 F-14050 Caen C\'edex, France
  \and Dipartimento di Fisica and INFN, Bologna, Italy } 
\date{Received: date / Revised version: date}
%
\abstract{ This contribution presents a review of our present
theoretical as well as experimental knowledge of different fluctuation
observables relevant to nuclear multifragmentation.
The possible connection between the presence of a fluctuation 
peak and the occurrence of a phase transition or a critical
phenomenon is critically analyzed. Many different phenomena can lead
both to the creation and to the suppression of a fluctuation peak. 
In particular, the role of constraints due to conservation laws 
and to data sorting is shown to be essential. 
From the experimental point of view, a comparison of the available 
fragmentation data reveals that there is a good agreement between 
different data sets of basic fluctuation observables, if the fragmenting
source is of comparable size. 
This compatibility suggests that the fragmentation process is largely
independent of the reaction mechanism (central versus peripheral collisions,
symmetric versus asymmetric systems, light ions versus heavy ion induced 
reactions).
Configurational energy fluctuations, that 
may give important information on the heat capacity of the fragmenting
system at the freeze out stage, are not fully compatible among different
data sets and require further analysis to properly account for Coulomb 
effects and secondary decays.
Some basic theoretical questions, concerning the interplay between 
the dynamics of the collision and the fragmentation process, and 
the cluster definition in dense and hot media, are still open and are
addressed at the end of the paper.  
A comparison with realistic models and/or
a quantitative analysis of the fluctuation properties will be 
needed to clarify in the next future the nature of the 
transition observed from compound nucleus evaporation 
to multi-fragment production. 
} 

\maketitle

\section{Fluctuations and phase transitions} 

Since the first inclusive heavy ion experiments, multifragmentation has been tentatively 
associated to a phase transition or a critical phenomenon. This expectation was triggered 
by the first pioneering theoretical studies of the nuclear phase diagram \cite{bertsch} 
which contains a coexistence region delimited, at each temperature below an upper critical value, by two critical 
points at different asymmetries\cite{mueller,ducoin}. 

Even more important, the first exclusive 
multifragmentation studies have shown that multifragmentation is a threshold process occurring
at a relatively well defined deposited energy\cite{nautilus,aladin,isis,tamain-here}. 
The wide variation of possible fragment partitions naturally leads to important fluctuations
of the associated partition sizes and energies.

Different observables have been proposed to measure such fluctuations. 
Using the general definition of the $n-th$ moment as
\begin{equation}
M_n = \sum_{Z_i\neq Z_{max}}Z_i^n \cdot n_i\left ( Z_i \right ),
\end{equation}
the variance of the charge distribution is measured by the second moment $M_2$
or by the normalized quantity\cite{perco}:
\begin{equation}
\gamma_2=\frac{M_2 M_0}{M_1^2}
\label{gamma2}
\end{equation}
The root mean square fluctuation per particle
\begin{equation}
\sigma_m=\sqrt{\langle \left ( Z_m/Z_0-\langle Z_m/Z_0 \rangle \right )^2\rangle}   
\label{rms}
\end{equation} 
of the distribution of the largest fragment $Z_m$ detected in each event
completes the information.
We will also consider the total fluctuation 
\begin{equation}
\Sigma^2_m= \langle Z_0 \rangle \sigma^2_m
\label{nvz}
\end{equation}
and  
the fluctuation 
\begin{equation}
\sigma^2_k= \langle \left ( E_p/A_0-\langle E_p/A_0 \rangle \right )^2\rangle
\label{sigmak}
\end{equation}
of the configurational 
energy per particle associated to each fragment partition $(k)$
\begin{equation}
E_p^{(k)}= \sum_{i=1}^{m_k} (BE)_i + \alpha^2 \sum_{i,j=1}^{m_k}  \frac{Z_i Z_j}{\langle |\vec{r}_i-\vec{r}_j|\rangle}
\end{equation}
where $m_k$ is the multiplicity of event $k$, $BE$ is the ground state 
binding energy of each fragment, and ${\langle |\vec{r}_i-\vec{r}_j|\rangle}$ is the average interfragment 
distance at the formation time.
The quantities $A_0$,$Z_0$ in eqs.(\ref{rms}),(\ref{sigmak}) represent the reconstructed
charge and mass of the fragmenting system, $Z_0=\sum_{i=1}^{m_k}Z_i$, 
$A_0=\sum_{i=1}^{m_k}A_i$.

In a simple statistical picture the fluctuation of any observable can be related to the 
associated generalized susceptibility by 
\begin{equation}
\chi = - \frac{\partial \langle A \rangle}{\partial \lambda}
= \langle A^2 \rangle -\langle A \rangle^2
\label{fluct-dis}
\end{equation}
where $\lambda$ is the intensive variable associated to the generic observable $A$.
Since the intensive variable associated to a particle density $N/V$ is the susceptibility 
$\chi=\partial \langle N \rangle /\partial \mu$,
then the large variance of the charge distribution observed in multifragmentation 
experiments could be connected to the diverging critical point fluctuation 
which would signal a diverging susceptibility and a diverging density correlation length.
The apparent self-similar behavior and scaling properties of fragment 
yields\cite{elliott-here} tends to support this intuitive picture.

\subsection{Finite size effects}
\begin{figure}[ht] 
\resizebox{0.70\columnwidth}{!}{\includegraphics{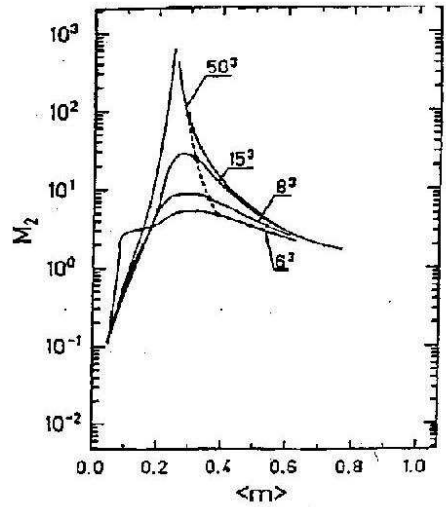}}
\caption{Second moment of the size distribution (see \cite{campi-here} for a precise definition)
as a function of the average cluster multiplicity for the three-dimensional percolation model
for different lattice sizes.   Figure is taken from \protect\cite{perco}. 
} 
\label{perco} 
\end{figure}

Many different effects can however blur this simple connection.
First of all, since fragmenting sources cannot exceed a few hundred nucleons, we have 
certainly to expect finite size rounding effects, which smooth the fluctuation signal\cite{perco}.
Not only the transition point is expected to be loosely defined and shifted in the finite
system as shown in the three-dimensional percolation model in Figure~\ref{perco}, but 
also the signal is qualitatively the same for a critical point, a first order transition
or even a continuous change or cross over.

Finite size effects have other consequences on the distribution than the simple smoothing of the transition.
It has been shown on different model calculations that the presence of conservation constraints
as well as the use of different event sorting procedures can sensibly distort the fluctuation observables.
To give a simple example, the presence of a peak in the largest fragment's size fluctuation as a function
of the energy deposit is trivially produced by the baryon number conservation constraint which forces this 
fluctuation to decrease with increasing average multiplicity\cite{elliott}. In the case of a genuine
critical behavior as for the percolation model, the fact of sorting events according to the percolation 
parameter $p$ or according to some other correlated observable, as for instance the total cluster multiplicity,
modifies\cite{elliott,aladin} the behavior of $m_2$, $\gamma_2$, and all other related moments\cite{campi-here} 
measuring the fluctuation properties of the system. 
All these effects can be understood in the general framework of the non- equivalence of statistical ensembles 
for finite systems, which we will discuss in the next chapter.

\subsection{Thermal invariance properties} 
\begin{figure}[ht]
\resizebox{0.75\columnwidth}{!}{\includegraphics{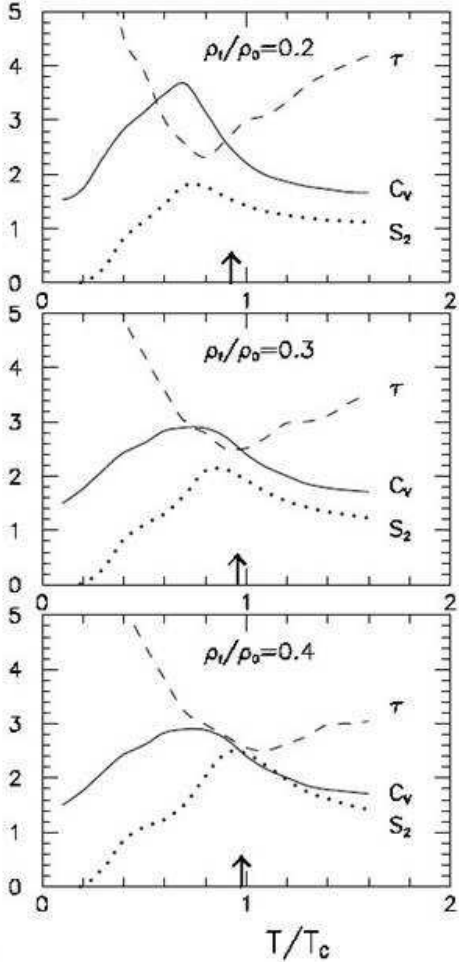}}
\caption{ Second moment of the charge ($S_2$) 
and of the energy ($C_v$) distribution as a function of temperature in the 
Lattice Gas model for different densities for a system of linear dimension L=7.  
Arrows: first order transition temperature in the thermodynamic limit. Figure taken from
ref.\protect\cite{dasgupta}. 
} 
\label{m2_rho} 
\end{figure}

Another problem when trying to connect a fluctuation peak to a phase transition or a critical behavior
in a finite system is given by the possible existence of thermodynamic ambiguities.
It has been observed by different independent works that in the framework of equilibrium fragmentation 
models the fluctuation behavior is qualitatively independent of the break up density 
\cite{raduta,richert,dasgupta,prl99}.
An example is given in Figure \ref{m2_rho}, which gives the second moment of the charge ($S_2=M_2-M_1^2$) 
and of the energy ($C_v$) distribution as a function of temperature in the Lattice Gas model 
for different break-up densities in the subcritical regime.

A peak in the fluctuation observables can 
be seen at all densities, at a temperature which is systematically below the critical temperature 
of the system and close to the first order transition temperature in the thermodynamic limit. 
A similar behavior has been observed in different fluctuation observables and also at supercritical 
densities along the Kertesz percolation line,
where the system does not present any phase transition. 
Table \ref{tab:1} gives, as a function of the lattice size, the inverse temperature
at which the variable $S_2$ shows a maximum in the three-dimensional 
IMFM model\cite{richert} at different densities.
%
\begin{table}
\caption{Inverse temperature at which the second moment $S_2=M_2-M_1^2$ 
is maximal for different
densities and lattice sizes in the three-dimensional 
IMFM model. Taken from ref.\cite{richert}}
\label{tab:1}       
\begin{tabular}{llll}
\hline\noalign{\smallskip}
L & $\beta_c(\rho=.3)$ & $\beta_c(\rho=.5)$ & $\beta_c(\rho=.7)$  \\
\noalign{\smallskip}\hline\noalign{\smallskip}
10 & .2560(5) & .225(3) &  .194(2) \\
16 & .2440(2) & .2230(5) & .1984(2) \\
20 & .23960(10) & .2227(4) & .1990(6) \\
24 & .2367(3) & .2227(2) & .2005(6) \\
\noalign{\smallskip}\hline
\end{tabular}
\end{table}
As a general statement, the fluctuation peak as well as the global scaling properties of the size 
distribution\cite{prl99,fisher} in these models can be found along a curve in the $T(\rho)$ diagram
passing through the thermodynamic critical point but extending in the subcritical as well as
supercritical region\cite{campi-kertesz}.
The subcritical behavior can be understood as a finite size effect, when the correlation
length, close to the first order transition point, becomes comparable to the linear size 
of the system, while the supercritical behavior is linked to the definition of clusters
in dense and hot media\cite{campi-kertesz}. For the subcritical region, a clusterization
algorithm has been suggested to eliminate such behaviors in Ising 
simulations\cite{elliott-here}.
The possible pertinence of all these observations
to experimental data is still a subject of debate, and essentially depends on the relationship
between the measured clusters and the cluster definitions of the models. 

Last but not least, the presence of different timescales in the reaction
\cite{bonasera-here,roy-here,dorso-here} 
and the dynamics of the fragmentation process may have important effects in the 
quantitative value of charge partition fluctuations\cite{claudio}, as we will discuss 
in the last chapter.

For all these reasons, it is clear that the well documented presence of a fluctuation peak
in the measured charge distributions\cite{tamain-here} cannot be taken as such 
as a proof of a critical behavior and/or phase transition. 
In order to connect the fluctuation behaviour to a phase transition and to conclude 
on its order it is indispensable to compare with models and/or to quantify the fluctuation peak.

\section{Theory} 
\subsection{Fluctuations and constraints}

\begin{figure}[ht] 
\resizebox{1.05\columnwidth}{!}{\includegraphics{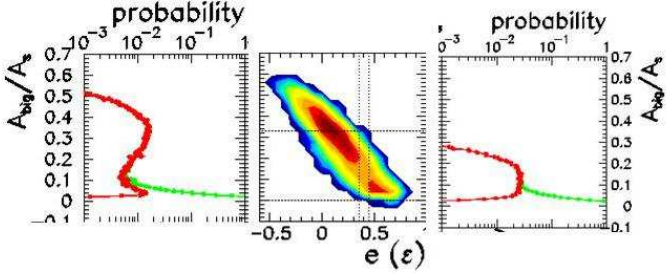}}
\caption{ Center: correlation between the largest fragment's size and the total 
energy in the isobar Lattice Gas model close to the transition temperature,
for a system of 216 particles.
Left side: projection over the $A_{big}$ direction. Right side: same as the left side,
but only events within a narrow energy interval around the average energy have been 
retained. 
} 
\label{constraint} 
\end{figure}

It is clear that fluctuations on a given observable $A$ will be suppressed if a constraint 
is applied to a variable correlated to $A$. This trivial fact has a deep 
thermodynamic meaning and is linked to the non equivalence of statistical ensembles in 
finite systems\cite{inequivalence}. Indeed the basic statistical relation between a fluctuation and the 
associated susceptibility eq.(\ref{fluct-dis}) is only valid in the ensemble in which 
the fluctuations of $A$ are such as to maximise the total entropy, under the constraint 
of $\langle A \rangle$ ("canonical" ensemble). The thermodynamics in the ensemble where 
the generic observable $A$ is controlled event by event ("microcanonical" ensemble), or
in the ensemble where $\sigma_A$ is externally fixed ("gaussian" ensemble\cite{challa})
is a perfectly defined statistical problem, but the thermodynamic relationships have to be explicitly 
worked out\cite{noi}. 
As an example we show in Figure \ref{constraint} the correlation between the size of the largest
cluster $A_{big}$ and the total energy in the isobar Lattice Gas model\cite{noi} at the transition temperature.
The presence of two energy solutions at the same temperature and pressure clearly shows that
the transition is in this case first order\cite{lopez-here}.  
The $A_{big}$ fluctuation properties are very different in the canonical ensemble (left part)
and in the microcanonical ensemble (right part) at the same (average) total energy. 
Because of the important correlation between the total energy and the fragmentation partition,
fragment size fluctuations can be compared only for samples with comparable widths of the 
energy distribution. 

From the experimental viewpoint, different constraints apply to fragmentation data 
and have to be taken into account. Apart from the sorting conditions\cite{tamain-here},
the collisional dynamics can also give important constraints to the fragmentation 
pattern ($e.g.$ flows, deformation in r-space and p-space).
This means that fluctuations have to be 
compared with calculations performed in the statistical 
ensemble corresponding to the pertinent experimental constraints\cite{botvina-here}.

\subsection{Fluctuations and susceptibilities}

\begin{figure}[ht] 
\resizebox{0.75\columnwidth}{!}{\includegraphics{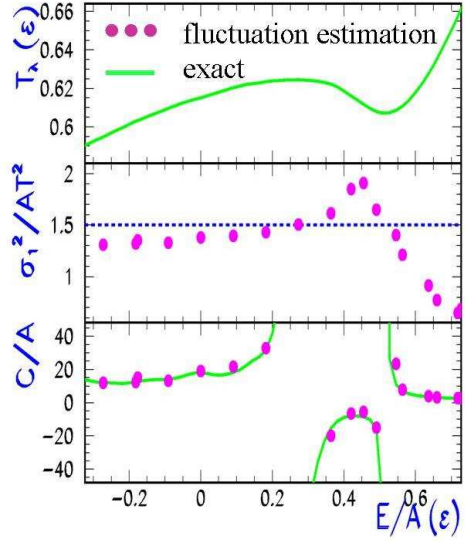}}
\caption{Temperature, normalized binding energy fluctuation and heat 
capacity in the microcanonical isobar Lattice Gas model as a function 
of the total energy for a system of 108 particles. In the lower panel the 
heat capacity estimated from fluctuations via eq.(\protect\ref{cneg})
(dots) is compared to the exact expression from the entropy curvature
(line).
Figure is taken from \protect\cite{europhys}. 
} 
\label{europhys} 
\end{figure}

In the last subsection we have stated that a connection between a fluctuation and the associated 
susceptibility can always be in principle worked out if the constraints acting on the observable
are known. 
In the case of sharp constraints ($e.g.$ fixed total mass, charge, deposited energy),
the connection between the fluctuation on a variable correlated to the constraint 
($e.g.$ size or charge of the largest fragment, configurational energy) and the associated 
susceptibility are in many cases analytical\cite{lebowitz,npa99,gros}.
If a conservation constraint $A=A_1+A_2=cst$ applies and the system can be splitted in two 
statistically independent components such that $W(A)=W(A_1)W(A_2)$, then 
the partial fluctuations are linked to the total susceptibility by 
\begin{equation}
\frac{\chi_1}{\chi}=1-\frac{\sigma^2_1}{\sigma^2_{ref}},
\end{equation}
where $\chi_i^{-1}=\partial^2_{A_i}W_i$, $\sigma^2_{ref}$ is the fluctuation of $A_1$
in the ensemble where only the average value $\langle A \rangle$ is constrained, and 
we have approximated the distribution of $A_1$ with a gaussian\cite{noi}.
The case of the total energy constraint has been particularly studied in the literature.
Indeed the total energy deposit can be (approximately\cite{viola-here}) measured 
event by event in $4\pi$ experiments, allowing to experimentally construct a microcanonical
ensemble by sorting. For classical systems with momentum independent interactions the potential
energy fluctuation $\sigma^2_I$ at a fixed total energy is linked to the total microcanonical 
heat capacity by
\begin{equation}
\frac{C_k}{C}=1-\frac{\sigma^2_k}{\sigma^2_{can}},
\label{cneg}
\end{equation}
where $C_k$,$C$ are the kinetic and total heat capacity, $\sigma^2_k=\sigma^2_I$ 
and $\sigma^2_{can}=c_k T^2$ is the kinetic energy fluctuation in the canonical ensemble.
Apart from the microstate equi-probability inherent to all statistical calculations, 
the above formula is obtained in the saddle point approximation for the partial 
energy distributions. The contribution of non gaussian tails can be also analytically
worked out\cite{npa99} and has been found to be negligible in all theoretical as well
as experimental data samples analyzed so far\cite{noi}.
An exemple of the quality of the approximation is given in Figure \ref{europhys}
which gives the temperature, normalized potential energy fluctuation and heat 
capacity in the isobar Lattice Gas model for a system of 108 particles.

\section{Experiment} 
 
\subsection{Effect of the sorting variable}
In this section we turn to compare different sets of experimental data available in the 
literature. A special attention has been paid by different collaborations to the largest fragment fluctuation 
$\sigma_m$ eq.(\ref{rms}) and to the $\gamma_2$
observable eq.(\ref{gamma2})\cite{aladin,ags1,ags2,ags3,eos,multics}. 
For all data sets of comparable total size these observables, as well as the others we will show in the next 
subsections, show a well defined peak at comparable values of the chosen 
sorting variable. This is an important and  
non trivial result considering that data are taken
with different apparata and the multifragmenting systems are obtained with very different 
reaction mechanisms.
The effect of the sorting variable is explored in table \ref{tab:sorting},
that gives the maximum value of $\gamma_2$ and $\sigma_m$ with different 
data sets sorted in bins 
of the total measured bound charge $Z_{bound}$, total measured charged particles
multiplicity $m$, or calorimetric excitation energy\cite{viola-here}.
\begin{table}
\caption{\small Maximum $\gamma_2$ and $\sigma_m$ values measured in the
break up of a $Au$ system 
within different data sets sorted in $Z_{bound}$,
total multiplicity ($m$) or calorimetric excitation energy ($\epsilon^*$).
Different values for the same case denote different bombarding energies.
Values taken from refs.\cite{aladin,elliott,ags1,ags2,ags3,eos,multics}}.
\label{tab:sorting}  
\begin{tabular}{lllllll}
\hline\noalign{\smallskip}
 & $\gamma_2$\protect\cite{aladin} & 
$\gamma_2$\protect\cite{ags1,ags2,ags3} 
& $\gamma_2$\protect\cite{eos,elliott} 
 & $\gamma_2$\protect\cite{multics} &
 $\sigma_m$\protect\cite{eos} 
&  $\sigma_m$\protect\cite{multics}\\
\noalign{\smallskip}\hline\noalign{\smallskip}
$Z_{bound}$ & 1.4 &  1.3 & & & & \\
m &  & 1.85& 3.2 & 2.23 & & .15 \\
 &  & 1.85 &  &  &  &  \\
  &  & 2.5 &  &  &  &  \\
$\epsilon^*$ & & & 3.7 & 2.5 & .12 & .14\\
\noalign{\smallskip}\hline
\end{tabular}
\end{table}
Even if the systematics should certainly be completed 
and errors should definitely be evaluated, 
we can observe from table \ref{tab:sorting} 
that different data sets show a reasonable agreement 
when the same sorting is employed.

We can also note that a higher $\gamma_2$ is systematically 
obtained when data are analyzed 
in bins of total charge multiplicity, with respect to a sorting in $Z_{bound}$.
This can be qualitatively understood if we recall that $\gamma_2$ 
measures the variance of the charge bound in fragments, 
and this quantity is obviously strongly correlated with $Z_{bound}$
and loosely correlated with $m$. 
The calorimetric excitation energy sorting leads to comparable 
results to the multiplicity sorting. The value of $\gamma_2$ is 
slightly increased, which may be explained by 
a reduced correlation of $\epsilon^*$ with respect to $m$ with the total 
fragment charge, since the excitation energy contains 
the extra information of the 
kinetic energy of the fragments. However the effect goes in the opposite
direction as the fluctuation of $Z_{m}$ is concerned.
A detailed study of the correlation coefficient between the considered 
observables and the sorting variables is needed to fully understand 
thess trends. 
It is also possible that the fluctuations obtained with these two sortings
may be compatible within error bars, which stresses the importance of 
an analysis of errors.
 
The fluctuation values appear to be largely independent of the reaction 
mechanism and incident energy\cite{aladin,ags1,ags3}. 
The only exception is the  value
$\gamma_2\approx 2.5$  obtained from emulsion data in ref.\cite{ags2}, which
is significantly higher than the values obtained at the other bombarding energies
for the same system. Such anomaly
might be due to the presence of fission events that have been excluded 
in the other analyses\cite{ags1,ags3}. The independence on the incident energy 
tends to show that the fragmentation process is 
essentially statistical.

\subsection{Effect of the system size}\label{size}
The effect of the system size is further analyzed in table \ref{tab:mass}.
All presented data are sorted in bins of calorimetric excitation energy.
%
\begin{table}
\caption{\small Maximum $\gamma_2$, $\Sigma^2_m$ and $\sigma_m$ values measured   
within different data sets for various system sizes $Z_0$. 
Different values for the same case denote different targets.
Values taken from refs.\cite{eos,nimrod,multics}}.
\label{tab:mass}     
\begin{tabular}{llll}
\hline\noalign{\smallskip}
$<Z_0>$ & $\gamma_2$\protect\cite{eos,nimrod,multics} &  
  $\Sigma^2_m$\protect\cite{eos,nimrod,multics} & 
 $\sigma_m$\protect\cite{eos,nimrod,multics}   
  \\
\noalign{\smallskip}\hline\noalign{\smallskip}
76 & 2.5  &   1.49 &  0.14     \\
59 & 3.7  &   0.85 &  0.12     \\
43 & 2.4  &   0.73 &  0.13    \\
27 & 1.75 &   0.39 &  0.125    \\
16 & 1.19 &   0.22 &  0.114 \\
   & 1.17 &   0.22 &  0.114 \\
   & 1.16 &   0.22 &  0.114 \\
\noalign{\smallskip}\hline
\end{tabular}
\end{table}

The fluctuation properties of quasi-projectile decay appear to be
largely independent of the target. This well known behavior at relativistic
energy\cite{aladin} appears confirmed in the case of the Nimrod 
experiment\cite{nimrod} which was performed with a beam energy
as low as 47 MeV/A. 
This suggests that a quasi-projectile emission source can be 
extracted\cite{tamain-here} in spite of the important midrapidity 
contribution in the Fermi energy regime\cite{roy-here}.

From table \ref{tab:mass} 
we can also see that $\Sigma^2_m$ 
decrease monotonically with the system mass. The evolution with the system size, at 
least in the size range analyzed, appears as a simple scaling behavior
as shown by the fact that the normalization to the source size in
$\sigma_m$ makes the fluctuation almost independent of the size. 
Similar conclusions can be drawn concerning the $\gamma_2$ observable, even
if the behavior for the heaviest sources is less clear.
This interesting scaling behavior should be confirmed using hyperscaling techniques\cite{campi-here}.

To conclude, we have seen that fluctuations can vary 
of a factor two changing the sorting variable. 
This stresses the need of confronting the experimental data with statistical predictions containing the same constraints, i.e. performed in the adapted statistical ensemble. Interesting enough, when the same sorting is adopted  
the different available data sets agree within $\approx$ 15\% both in the value of the 
peak and in the position where the peak is observed. 
More data are needed to confirm these trends.

\subsection{Configurational energy fluctuations}
One of the most interesting aspects of studying fluctuation observables, is their possible connection
with a susceptibility or a heat capacity via eq.(\ref{cneg}).
Configurational energy fluctuations have been studied at length by the Multics 
collaboration\cite{plb,palluto,mich-last} and by the Indra collaboration\cite{palluto,nicolas,lopez,bernard}
on Au sources. The observable used in these studies is an estimation of the energy stored 
in the configurational degrees of freedom at the time of fragment formation, defined as follows
\begin{eqnarray}
E_I&=&\sum_{i=1}^{N_{imf}}Q\left ( Z_i^p,A_i^p \right )\nonumber \\
&+& \sum_{i=n,p,d,t,^3He,^4He} 
Q\left ( Z_i,A_i \right )M_i^p \\
&+& V_{coul}\left ( \left \{ Z_i^p\right \}, V_{FO} \right )\nonumber
\end{eqnarray}
where $Q$ indicates the mass defects and $V_{coul}$ the coulomb energy. 
The measured fragment charges $Z_i$ and lcp multiplicities $M_i$ are corrected in each event 
to approximately account for secondary decay 
\begin{eqnarray}
Z_i^p&=&Z_i+\langle M^{ev}_H+2M^{ev}_{He} \rangle \frac {Z_i}{\sum_{i=1}^{N_{imf}}Z_i} \\
M_i^p&=&M_i-\langle M^{ev}_i \rangle
\end{eqnarray}
where $\langle M^{ev}_i \rangle $ is the estimated multiplicity of secondary emitted light charged particles
for each calorimetric excitation energy bin. 

\begin{figure}[ht] 
\resizebox{1.05\columnwidth}{!}{\includegraphics{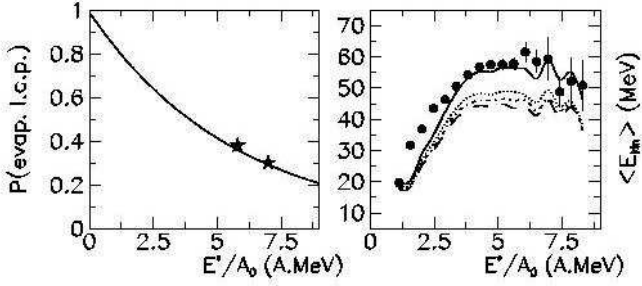}}
\caption{  Left part: percentage of secondarily emitted light charged particles taken 
from correlation functions measurements (see ref.\cite{giuseppe-here}).
Right part: total measured fragment kinetic energy (points) compared with Coulomb trajectories 
calculations where the volume is changed from $6V_0$ to $3V_0$.
Both quantities are plotted as a function of the calorimetric excitation energy.
Figure is taken from \protect\cite{palluto}. 
} 
\label{palluto} 
\end{figure} 

Three quantities need to be estimated in each excitation energy bin to compute $E_I$ 
\begin{enumerate}
	\item The freeze out volume $V_{FO}$ which determines the total Coulomb energy. Its average value
	 is deduced from the measured fragment kinetic energies through Coulomb trajectories calculations
	 (see Figure \ref{palluto}, right part). 
	 \item the average multiplicities of secondarily emitted particles $\langle M^{ev}_{lcp}\rangle$ to account for 
	  side feeding effects. They are deduced from fragment-particle correlation functions (see Figure \ref{palluto}, left part).
	\item the isotopic content $A_i^p/Z_i^p$ of primary fragments. It is assumed that it is equal to the isotopic content of the   
	        fragmenting system. This quantity allows in turn to determine the number of free neutrons at freeze out from baryon 
	        number conservation.
\end{enumerate}

\begin{figure}[ht] 
\resizebox{1.\columnwidth}{!}{\includegraphics{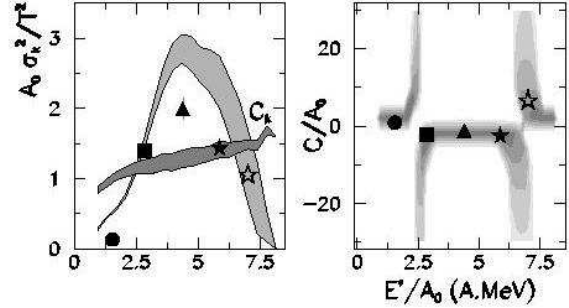}}
\caption{  Left side: normalized fluctuation of $E_I$ and estimated $C_k$ (see text) as a function of the 
calorimetric excitation energy. Grey zone: peripheral 35A.MeV Au+Au collisions. Symbols: central Au+C, 
Au+Cu, Au+Au at 25 and 35 A.MeV. Right side: heat capacity from eq.\ref{cneg}.
Figure is taken from \protect\cite{mich-last}. 
} 
\label{cneg_exp} 
\end{figure}

A general protocol has been proposed to minimize the spurious fluctuations due to the implementation 
of this missing information\cite{palluto}.
The resulting fluctuation of $E_I$ $\sigma^2_I=\sigma^2_k$ 
is shown for different Multics data in figure \ref{cneg_exp}.
The temperature has been estimated alternatively using isotopic thermometers or solving the kinetic
equation of state and comes out to be in good agreement\cite{mich-last} with the general temperature
systematics \cite{natowiz-here} (around 4.5 MeV in the fragmentation region). 
Similar to the other fluctuation observables, 
configurational energy fluctuations show a well pronounced peak 
at an excitation energy around 5 A.MeV. 
This general feature is apparent in Multics\cite{mich-last}, Indra\cite{bernard},
Isis\cite{lefort} and Nimrod\cite{nimrod} data. 
The only exception is EOS data\cite{srivastava} where this fluctuation appears
monotonically decreasing.

In the hypothesis of thermal equilibrium at the freeze out configuration 
this fluctuation is a measurement of the heat capacity according to eq.(\ref{cneg}).
The value expected for this fluctuation in the canonical ensemble
can be written as $\sigma^2_{can}=c_kT^2$. The kinetic heat capacity $c_k$ is calculated from the measured 
fragment yields\cite{palluto}. We can see that the fluctuation peak 
overcomes the upper classical limit $c_k=3/2$ suggesting a negative heat capacity as expected in a first 
order phase transition analyzed in the microcanonical ensemble\cite{chomaz-here,gross-here}.    

\begin{figure}[ht] 
\resizebox{1.\columnwidth}{!}{\includegraphics{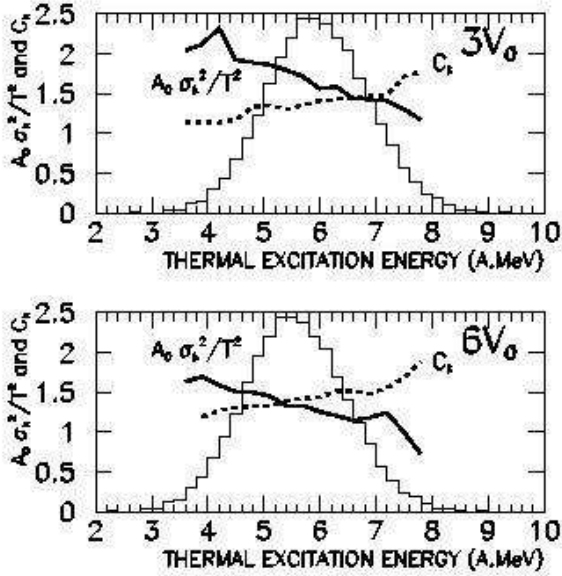}}
\caption{Normalized fluctuation and kinetic heat capacity (stars) for 32 A.MeV Xe+Sn
central collisions measured by the INDRA collaboration as a function of 
the calorimetric excitation energy with two different hypothesis on the 
freeze out volume. The histogram gives the event distribution.
Figure is taken from \protect\cite{nicolas,palluto}. } 
\label{xesn} 
\end{figure}

The same analysis performed on Indra data of central Xe+Sn collisions at different bombarding energies 
leads to compatible temperatures and volumes and a fluctuation estimation that agrees within 25\% with 
the presented Multics results\cite{palluto}, as shown for the 32 A.MeV data in figure \ref{xesn} (upper part). 
In the absence of isotopic resolution for fragments, Coulomb repulsion cannot be distinguished
from a radial collective expansion due to a possible initial compression. 
If an important radial flow component is assumed for these 
central collisions, data can also be compatible with a bigger freeze out volume (lower part of the figure)
leading to a shift of the abnormal fluctuation behavior towards lower energy.
This volume/flow ambiguity in central collisions 
can only be solved with third generation multidetectors\cite{leneindre-here}.

\begin{figure}[ht] 
\resizebox{1.05\columnwidth}{!}{\includegraphics{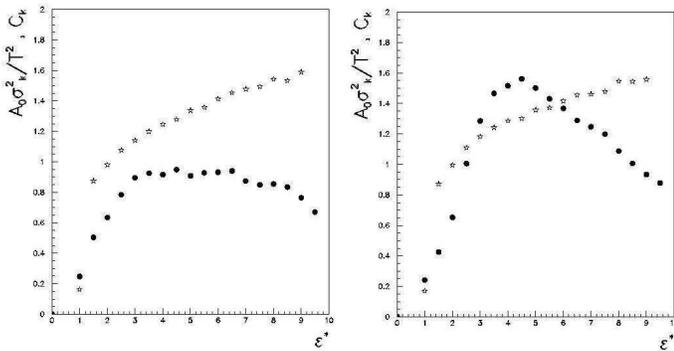}}
\caption{ Normalized fluctuation and kinetic heat capacity (stars) for 80 A.MeV Au+Au
peripheral collisions measured by the INDRA and ALADIN collaboration as a function of 
the calorimetric excitation energy, for all quasi-projectile events (left side) and 
after subtraction of events elongated along the beam axis (right side).   
Figure is taken from \protect\cite{bernard}. } 
\label{cneg_au} 
\end{figure}

Indra data on the same Au quasi-projectile analyzed
by the Multics collaboration lead to a fluctuation measurement about 40\% lower, see figure \ref{cneg_au}. 
This difference is tentatively
explained as an effect of emission from the neck which leads to a reduced occupation of the available
phase space\cite{bernard}.  

\begin{figure}[ht] 
\resizebox{0.75\columnwidth}{!}{\includegraphics{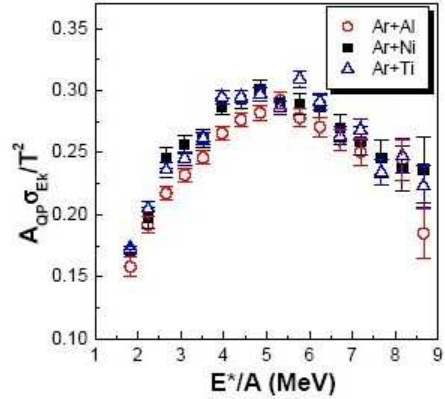}}
\caption{ Normalized fluctuation and kinetic heat capacity for 47 A.MeV Argon quasi-projectiles
on different targets measured by the NIMROD collaboration as a function of 
the calorimetric excitation energy. Figure is taken from \protect\cite{nimrod}. } 
\label{ma} 
\end{figure}

Recent Nimrod data\cite{nimrod} on the fragmentation of a much lighter system 
show a similar value for the energy corresponding to the fluctuation peak, but
fluctuation a factor 10 lower than for Multics data, as shown in figure \ref{ma}. 
If we consider the global fluctuation $<A_0>\sigma^2_k$ without the normalization
to the estimated temperature, this factor is reduced to about a factor 4.
These results go in the same direction as the general behavior
of $\Sigma^2_m$ that we have analyzed in section \ref{size}. Recall that the 
fluctuation of the biggest fragment for the quasi-Au source\cite{multics}
is a factor 6.8 higher than for the quasi-Ar\cite{nimrod}.
This fluctuation reduction seems then to be a general feature 
of light systems fragmentation and has been
tentatively explained as an effect of the higher temperature that light systems can sustain\cite{nimrod}. In this interpretation, a higher temperature region of 
the phase diagram, possibly above the critical point, 
is explored in the fragmentation of light systems, and 
the first order phase transition observed in heavy nuclei becomes a smooth cross over.

As a general remark, the configurational energy fluctuation signal 
is a very interesting one due to its
possible connection with a heat capacity, but it is also a very indirect and fragile 
experimental signal which needs precise calorimetric measurements, a careful data analysis, 
extensive simulations to assess the effect of the different hypotheses in the 
event sorting and reconstruction procedure. Moreover the different techniques 
to exclude or minimize preequilibrium and neck
emission seem to have a strong influence in the absolute value of fluctuations.

The evaluation of systematic errors in fluctuation measurements is necessary to achieve a
quantitative estimation of fluctuations: some first encouraging results in this direction have
been presented in ref.\cite{mich-last}.    
The confirmation (or infirmation) of the fluctuation enhancement is certainly 
one of the most important challenges of the field in the next years 
with third generations multidetectors.

\section{Open questions} 

The possibility of accessing a thermodynamic information 
on the nuclear phase diagram from measured fragment properties 
entirely relies on the representation of the system at the freeze out 
stage as an ideal gas of fragments\cite{botvina-here} in thermal equilibrium.
This is true for fluctuation observables as well as for all other thermodynamic
analyses\cite{natowiz-here,elliott-here}. This is an important conceptual 
point which is presently largely debated in the heavy ion community.

\begin{figure}[ht] 
\resizebox{0.75\columnwidth}{!}{\includegraphics{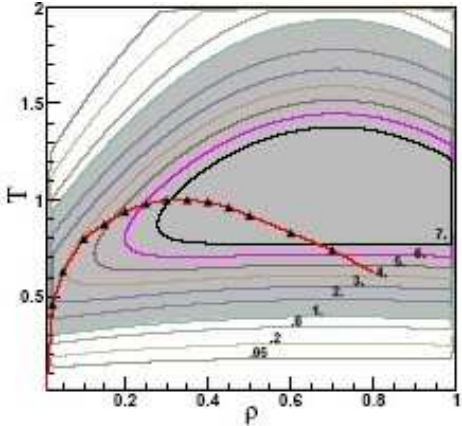}}
\caption{ Phase diagram of 64 Lennard Jones particles confined in a box. 
Filled triangles give the coexistence border. Isocontours give values
of normalized fluctuations $\sigma^2_k/\sigma^2_{can}$ 
calculated from the ground state Q value of 
clusters defined with the Hill algorithm. 
Figure is taken from \protect\cite{campi}. 
} 
\label{campi} 
\end{figure}

\begin{figure}[ht] 
\resizebox{0.95\columnwidth}{!}{\includegraphics{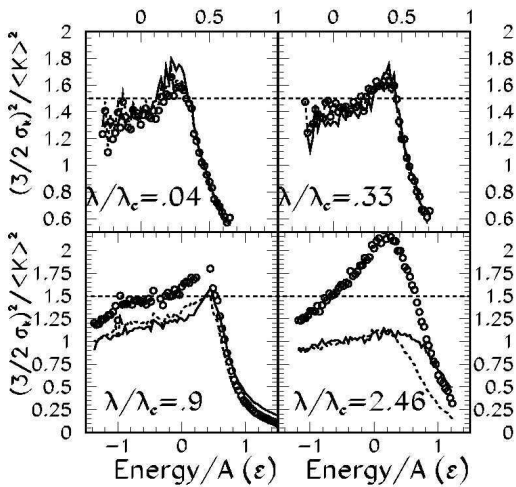}}
\caption{Normalized fluctuations $\sigma^2_k/T^2$ as a function of energy 
for a system of 216 Lattice Gas particles in the isobar ensemble 
at different pressures. Full lines: exact results. Symbols: estimation 
from the ground state Q value of Coniglio-Klein clusters. Dashed lines:
as the symbols, but data are sampled in bins of energy reconstructed
from cluster kinetic energies and sizes. $\lambda_c$ gives the critical pressure.
Figure is taken from \protect\cite{tracking}. 
} 
\label{tracking} 
\end{figure}

A first open question concerns the structure of the systems at the freeze out
stage, i.e. at the time when fragments decouple from each other. Contrary to 
the ultrarelativistic regime\cite{randrup-here}, we do not expect much difference 
between the chemical and kinetic decoupling times due to the small collective 
motions implied in these low energy collisions. We can therefore speak at least in a first 
approximation of a single freeze out time. If at this time the system is still
relatively dense, the cluster properties may be very different from the ones
asymptotically measured, and the question arises\cite{campi-kertesz} whether the 
energetic information measured on ground state properties can be taken backward 
in time up to the freeze out.
Calculations from classical molecular dynamics\cite{campi} show that the ground state 
Q-value is a very bad approximation of the interaction energy of Hill clusters in 
dense systems. This is due both to the deformation of clusters when recognized in a dense 
medium through the Hill algorithm, and to the interaction energy among clusters in dense 
configurations where clusters surfaces touch. 
As a consequence, comparable fluctuations are obtained in the subcritical
and supercritical region of the Lennard-Jones phase diagram. This result is shown in 
fig.\ref{campi}.  
Calculations in a similar model, the Lattice Gas model,
show that even in the supercritical regime the correct 
fluctuation behavior can be obtained if both the total energy 
and the interaction energy are consistently estimated with the 
same approximate algorithm as it is done in the experimental 
data analysis\cite{tracking}. Indeed the high value of the estimated 
configurational energy $Q$ fluctuations is essentially 
due to the spurious fluctuation of the total energy $E_K+E_I$
obtained when $E_I$ is estimated through $Q$; such an effect is 
eliminated if data are analyzed in bins of $E_K+Q$. 
This calculation is shown in fig.\ref{tracking}. 

\begin{figure}[ht] 
\resizebox{1.\columnwidth}{!}{\includegraphics{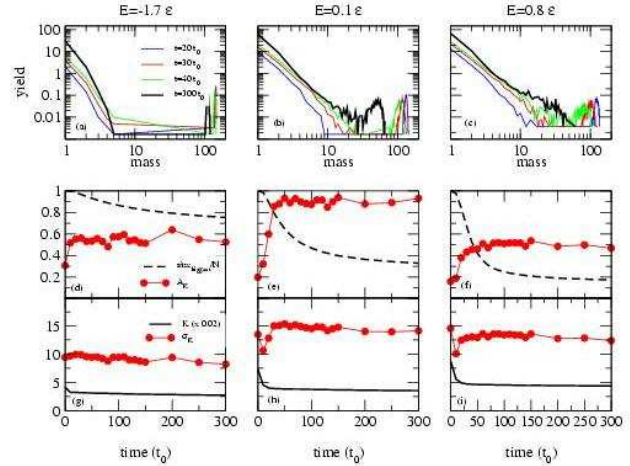}}
\caption{Time evolution of a Lennard Jones system initially confined 
in a dense supercritical configuration and freely expanding in the vacuum
at different total energies. Upper part: MST fragment size distribution at 
different times. Lower part: average kinetic energy (full lines), total (lower 
symbols) and normalized (upper symbols) kinetic 
energy fluctuations, and size of the largest MST cluster (dashed lines).
Abnormal fluctuations in these units correspond to $A_k>\approx 0.7$.
  Figure is taken from \protect\cite{ariel}. 
} 
\label{cherno} 
\end{figure}

A second related question which needs further work is the
relevance of the equilibrium assumption at freeze out. 
Molecular dynamics models applied to study the time evolution
of the reaction\cite{claudio,ariel,akira,aichelin} predict 
that the decoupling between fragment degrees of freedom 
(freeze out) occurs very rapidly during the reaction.
At this stage however the configuration is considerably
diluted due to the early presence of collective 
motions\cite{claudio,dorso-here}.
An example taken from classical molecular dynamics 
for an initially equilibrated compact configuration 
freely evolving in the vacuum, is shown in fig.\ref{cherno}.
At this reaction stage cluster energies may be well approximated 
(within a side feeding correction) by their asymptotically 
measured values, but it is not clear whether this 
configuration can correspond to an equilibrium, 
more precisely whether the hypothesis 
of equiprobability of the different charge partitions
holds.  

\section{Conclusions and outlooks} 
In this paper we have presented a short review 
of the experimental as well as theoretical studies of fluctuation
observables of fragments produced in a 
multifragmentation heavy ion reaction. 
The aim of these studies 
is the understanding of the nature of the nuclear fragmentation
transition as well as the
thermodynamic characterization of the finite temperature 
nuclear phase diagram. 
This vast and ambitious program is still in its infancy.
Many promising results already exist, but the analyses are 
not yet conclusive and need to be 
intensively pursued in the future.

The nuclear fragmentation phenomenon, 
well documented by a series of independent experiments\cite{tamain-here},
presents many features compatible with a critical phenomenon\cite{campi-here}
or a phase transition\cite{elliott-here,lopez-here}. Only a careful study 
of fluctuation properties will allow to discriminate between 
the different scenarii. Even more important, the phase diagram 
of finite nuclei is theoretically expected to present an anomalous 
thermodynamics\cite{gross-here,chomaz-here} which should be 
characteristic of any non extensive system undergoing first order 
phase transitions in the thermodynamic limit. Once the difficulties
linked to the imperfect detection and sorting ambiguities will be
overcome, fluctuation observables will be a unique tool to 
quantitatively study this new thermodynamics with its interdisciplinary 
applications\cite{gross-here,chomaz-here,pleimling}.

From the theoretical point of view, the theoretical connections between 
fluctuations and susceptibilities in the different statistical ensembles 
are well established, and the different experimental contraints can be 
consistently adressed by the theory. However, the evaluation of a thermodynamics
for a clusterized system opens the difficult theoretical problem of 
cluster definition in dense quantum media. To produce quantitative 
estimations of measurable fluctuation observables the pertinence of classical
models has to be checked through detailed comparisons with microscopic\cite{akira}
and macroscopic\cite{botvina-here} nuclear models.

On the experimental side, multiplicities and size fluctuations agree reasonably well
if comparable size fragmenting systems are studied, even if the effect of the system
size has to be clarified. Configurational energy fluctuations are especially 
interesting because of their possible connection with a heat capacity measurement.
The methodology to extract such fluctuations from fragmentation data is presently 
under debate, in particular a careful analysis of systematic errors is presently
undertaken\cite{mich-last}. From a more conceptual point of view, the influence of 
the different time scales in the reaction dynamics has to be clarified. 
Configurational energy fluctuations may be subject to strong ambiguities since they use 
information from all the particles of the event, and this information is integrated
over the whole reaction dynamics. In this respect, an interesting complementary observable 
may be given by fluctuation of the heaviest cluster size\cite{lopez-here,gros}.
 
To solve the existing ambiguities we need full comparisons with a well defined protocol  
and consistency checks between different data sets. 
The simultaneous measurement of fragments  mass and charge on a $4\pi$ geometry\cite{leneindre-here} will be 
essential to measure the basic variable of any thermodynamic study, namely the deposited energy. 
No definitive conclusion about the occurrence of a thermodynamic phase transition and its order 
can be drawn without this detection upgrade.  
%
 

\begin{thebibliography}{99}

\bibitem{bertsch} G.Bertsch, P.J.Siemens, Phys. Lett. B 126 (1983)9. 
\bibitem{mueller}  H.Muller and B.Serot, Phys. Rev. C52 (1995) 2072.  
\bibitem{ducoin} C.Ducoin et al, nucl-th/0512029.
\bibitem{nautilus}	G.Bizard et al, Physics Letters B 208 (1993) 162.
\bibitem{aladin}	A.Schuttauf et al, Nuclear Physics A 607 (1996) 457.
\bibitem{isis}	T.Beaulieu et al, Physical Review C 64 (2001) 064604.
\bibitem{tamain-here} B.Tamain, contribution to this book.
\bibitem{perco}	X.Campi, Physics Letters B 208 (1988) 351.
\bibitem{elliott-here} J.B.Elliott et al, contribution to this book.
\bibitem{elliott}	J.Elliott et al, Physical Review C 62 (2000) 064603.
\bibitem{campi-here} Y.G.Ma, contribution to this book.
\bibitem{raduta}Al.Raduta et al,  Phys.Rev. C 65(2002)034606.
\bibitem{richert}	J.Carmona et al, Nuclear Physics A 643 (1998) 115.
\bibitem{dasgupta}	J.Pan et al, Physical Review Letters 80 (1998) 1182.
\bibitem{prl99}	Ph.Chomaz, F.Gulminelli, Physical Review Letters 82 (1999) 1402.
\bibitem{fisher} F.Gulminelli et al, Phys.Rev.C 65(2002) 051601.
\bibitem{campi-kertesz} N.Sator, Phys.Rep. 376 (2003) 1.
\bibitem{bonasera-here} A.Bonasera et al, contribution to this book.
\bibitem{roy-here} M.Di Toro et al, contribution to this book.
\bibitem{dorso-here} C.Dorso, contribution to this book.
\bibitem{claudio}	P.Balenzuela et al, Physical Review C 66 (2002) 024613.
\bibitem{inequivalence}	F.Gulminelli et al, Physical Review E 68 (2003) 026120.
\bibitem{challa} M.S.Challa, J.H.Hetherington, Phys.Rev.Lett. 60 (1988) 77.
\bibitem{noi} F.Gulminelli, Ann. Phys. Fr. 29 (2004)6.  
\bibitem{lopez-here} O.Lopez and M.F.Rivet, contribution to this book.
\bibitem{botvina-here} A.Botvina, contribution to this book.
\bibitem{lebowitz}	J.L.Lebowitz et al, Physical Review 153 (1967) 250.
\bibitem{npa99}	F.Gulminelli,Ph.Chomaz ,Nuclear Physics A 647 (1999) 153.
\bibitem{gros} F.Gulminelli,Ph.Chomaz, Physical Review C 71 (2005) 054607.  
\bibitem{viola-here} V.Viola and R.Bougault, contribution to this book.
\bibitem{europhys} F.Gulminelli, Ph.Chomaz and V.Duflot, Europhys.Lett.50 (2000) 434.	
\bibitem{ags1}	P.L.Jain et al, Physical Review C 50 (1994) 1085.
\bibitem{ags2}	M.I.Adamovitch et al, European Journal of Physics A 1 (1998) 77.
\bibitem{ags3}	D.Kudzia et al, Physical Review C 68 (2003) 054903.
\bibitem{eos} J.B.Elliott et al, Physical Review C 67 (2003) 024609.
\bibitem{multics}	A.Bonasera et al, La Rivista del Nuovo Cimento 23 (2000) 1; M.D'Agostino, private communication.
\bibitem{nimrod}	Y.G.Ma et al, Nucl.Phys. A 749 (2005) 106; Y.G.Ma, private communication.   
\bibitem{plb} M.D'Agostino et al, Physics Letters B 473 (2000) 219.
\bibitem{palluto}	M.D'Agostino et al, Nuclear Physics A 699 (2002) 795.
\bibitem{mich-last}	M.D'Agostino et al, Nuclear Physics A 734 (2004) 512.
\bibitem{nicolas}	N.Leneindre, PhD, http://tel.ccsd.cnrs.fr/tel-0003741
\bibitem{lopez}	O.Lopez et al, proceedings of the IWM meeting, Caen, 2003
\bibitem{bernard} M.Pichon et al, nucl-ex/0602003 and PhD, http://tel.ccsd.cnrs.fr/tel-0007451.
\bibitem{natowiz-here} J.Natowiz and K.H.Schmidt, contribution to this book.
\bibitem{lefort} T.Lefort and V.Viola, private communication. 
\bibitem{srivastava} B.K.Srivastava et al., Phys.Rev. C65 (2002) 054617. 
\bibitem{giuseppe-here} G.Verde et al, contribution to this book.
\bibitem{chomaz-here} P.Chomaz, contribution to this book.
\bibitem{gross-here} D.Gross, contribution to this book.
\bibitem{randrup-here} J.Randrup and I.Mishustin, contribution to this book.
\bibitem{campi}	 X.Campi et al, Phys.Rev.C 71 (2005) 41601.
\bibitem{tracking}	F.Gulminelli, Ph.Chomaz, M.D'Agostino,Phys.Rev. C72 (2005) 064618.
\bibitem{ariel}	A.Chernomoretz et al, Physical Review C 69 (2004) 034610.
\bibitem{akira}  A.Ono, H.Horiuchi, Progr. Part. Nucl. Phys. 53 (2004) 501.
\bibitem{aichelin} A.D.Sood, R.K.Puri, J.Aichelin, Phys.Lett. 594 (2004) 260.  
\bibitem{pleimling} H.Behringer et al, J.Phys.A 38 (2005) 973. 
\bibitem{leneindre-here} N.Leneindre et al., contribution to this book.

\end{thebibliography}

\end{document}